\documentclass[iop, apjl, tighten]{emulateapj}

\usepackage{apjfonts}
\usepackage{amsmath}
\usepackage{amssymb}
\usepackage{mathrsfs}
\usepackage{bm}
\usepackage[backref,breaklinks,colorlinks,citecolor=blue]{hyperref}
\usepackage[all]{hypcap}

\usepackage{enumitem}
\usepackage{booktabs}
\usepackage[normalem]{ulem}

\begin{document}

\title{Absolute Calibration of Diffuse Radio Surveys at $45$ and $150$~MH\lowercase{z}} %
\author{
Raul A. Monsalve\altaffilmark{1,2,3},
Alan E. E. Rogers\altaffilmark{4},
Judd D. Bowman\altaffilmark{2},
Nivedita Mahesh\altaffilmark{2},
Steven G. Murray\altaffilmark{2},
Thomas J. Mozdzen\altaffilmark{2},
Leroy Johnson\altaffilmark{2},
John Barrett\altaffilmark{4},
Titu Samson\altaffilmark{2},
and David Lewis\altaffilmark{2}
}

\affil{$^1$Department of Physics and McGill Space Institute, McGill University, Montr\'eal, QC H3A 2T8, Canada; \href{mailto:Raul.Monsalve@mcgill.ca}{Raul.Monsalve@mcgill.ca}}
\affil{$^2$School of Earth and Space Exploration, Arizona State University, AZ 85287, USA}
\affil{$^3$Facultad de Ingenier\'ia, Universidad Cat\'olica de la Sant\'isima Concepci\'on, Alonso de Ribera 2850, Concepci\'on, Chile}
\affil{$^4$Haystack Observatory, Massachusetts Institute of Technology, MA 01886, USA}

\begin{abstract}

We use EDGES measurements to determine scale and zero-level corrections to the diffuse radio surveys by Guzm\'an et al. at $45$~MHz and Landecker \& Wielebinski at $150$~MHz. We find that the Guzm\'an et al. map requires a scale correction of $1.076 \pm 0.034$~($2\sigma$) and a zero-level correction of $-160 \pm 78$~K~($2\sigma$) to best-fit the EDGES data. For the Landecker \& Wielebinski map, the scale correction is $1.112 \pm 0.023$~($2\sigma$) and the zero-level correction is $0.7 \pm 6.0$~K~($2\sigma$). The correction uncertainties are dominated by systematic effects, of which the most significant are uncertainty in the calibration of the EDGES receivers, antenna pointing, and tropospheric and ionospheric effects. We propagate the correction uncertainties to estimate the uncertainties in the corrected maps themselves and find that the $2\sigma$ uncertainty in the map brightness temperature is in the range $3.2-7.5\%$ for the Guzm\'an et al. map and $2.1-9.0\%$ for the Landecker \& Wielebinski map, with the largest percent uncertainties occurring at high Galactic latitudes. The corrected maps could be used to improve existing diffuse low-frequency radio sky models, which are essential tools in analyses of cosmological $21$~cm observations, as well as to investigate the existence of a radio monopole excess above the cosmic microwave background and known Galactic and extragalactic contributions.

\end{abstract}


\section{Introduction}
\label{section_introduction}

Accurate knowledge of the intensity and structure of the diffuse radio sky below $200$~MHz is critical for studies of the neutral hydrogen $21$~cm signal from the dark ages, cosmic dawn, and epoch of reionization \citep{madau1997, shaver1999, tozzi2000}. At these frequencies the diffuse radio sky is dominated by Galactic and extragalactic synchrotron radiation, which acts as a foreground to the cosmological $21$~cm signal. The brightness temperature of this foreground is between hundreds and tens of thousands of Kelvin \citep{dowell2017, mozdzen2017}, while the $21$~cm signal is expected to be at least four orders of magnitude smaller \citep{furlanetto2006}. Estimating and removing the contribution from the diffuse radio foreground with high accuracy represents one of the biggest challenges in $21$~cm cosmology \citep[e.g.,][]{nhan2019}. 

Predictions for the diffuse radio sky at arbitrary frequencies can be computed using numerical codes such as \verb|GALPROP| \citep{strong2011, orlando2013} and \verb|hammurabi| \citep{waelkens2009, wang2020}. Among other products, these codes generate synchrotron radiation maps from models of cosmic ray propagation and the Galactic magnetic field. Predictions for the diffuse radio sky can also be computed from models based on publicly available sky surveys. These models are widely used in $21$~cm cosmology analyses and include: (1) the Global Sky Model (GSM) \citep{de_oliveira_costa2008}; (2) the Low Frequency Sky Model (LFSM) \citep{dowell2017}; (3) the improved GSM \citep{zheng2017}; and (4) the Global Model for the Radio Sky Spectrum (GMOSS) \citep{rao2017}. These models can yield all-sky maps at arbitrary frequencies by interpolating between the frequencies of the surveys. In particular, the LFSM relies on the surveys from the literature used by the other models but also on recent measurements with the Long-Wavelength Array~1 (LWA1) at $40-80$~MHz, which have coverage in declination down to $\delta=-40^{\circ}$. The maps generated by these models have an accuracy that is limited, to first order, by the uncertainties in the brightness temperature scale and the zero level of the surveys, which typically are between a few and tens of percent and at least a few Kelvin, respectively.

Accurate measurements of the diffuse radio sky, especially at high Galactic latitudes, have also become a priority in light of the report by ARCADE~2 \citep{fixsen2011, seiffert2011, singal2018} and LWA1 \citep{dowell2018} of a monopole `excess' above the cosmic microwave background (CMB) and known Galactic and extragalactic contributions. Several possible origins for this excess have been suggested \citep[e.g.,][]{fornengo2011}, including radio emission during or before the cosmic dawn \citep[e.g.,][]{cline2013, biermann2014}. Recent studies have discussed how a radio background in the early universe above the CMB, in addition to potentially explaining the reported excess, could have an impact on the cosmological $21$~cm signal \citep[e.g.,][]{feng2018, ewallwice2018, mirocha2018, fialkov2019, mondal2020, reis2020, caputo2020}. Although the claim of a monopole excess was disputed by \citet{subrahmanyan2013}, who conclude that it corresponds to Galactic synchrotron, it was supported by the diffuse component separation analysis of \citet{fornengo2014} and by the analysis in \citet{vasilenko2017} of measurements at $14.7-25$~MHz over a low-emission region of the northern sky with the UTR-2 array \citep{braude1978, vasilenko2006}. Thus, the existence of a monopole excess has not been fully resolved and its confirmation or disconfirmation could potentially lead to constraints on the high-redshift universe. Because the results by ARCADE~2 were obtained from a combined analysis of their own measurements at $3-90$~GHz, the CMB measurement by FIRAS \citep{mather1999}, and publicly available radio surveys, any corrections made to the surveys will have an impact on the estimate for the excess. 

The survey that has been subject to the largest number of refinements, in the form of destriping, filtering, and zero-level correction, is the Haslam $408$-MHz map \citep{haslam1981, haslam1982, bennett2003, remazeilles2015, wehus2017}. The need for these refinements mainly arises from its widespread use as a synchrotron template in studies of the CMB, $21$~cm cosmology, intensity mapping, and the interstellar medium \citep[e.g.,][]{strong2011,vedantham2014,planck2016,lattanzi2017,smoot2017,dickinson2019,planck2020}. It is necessary to determine similar corrections for the surveys at lower frequencies in order to improve our understanding of the radio sky and increase the accuracy achieved by analyses of the high-redshift $21$~cm signal and the radio monopole.

Here we report coordinate-independent corrections to the brightness temperature scale and zero-level of the diffuse radio maps from \citet{guzman2011} at $45$~MHz (henceforth G45 map) and \citet{landecker1970} at $150$~MHz (henceforth LW150 map). To determine the corrections, we use antenna temperature measurements conducted with four implementations of the Experiment to Detect the Global EoR Signature (EDGES) \citep{bowman2018}. The measurements were done from the Murchison Radioastronomy Observatory (MRO) at a latitude of $-26.7^{\circ}$. Although the EDGES instruments observe the sky across a wide frequency range with the objective of detecting the sky-averaged redshifted $21$~cm signal from the cosmic dawn and the epoch of reionization \citep{monsalve2017b, bowman2018}, in this analysis we only use data at the same frequency as the G45 and LW150 maps, i.e., $45$ and $150$~MHz. The observations span almost $24$~h of local sidereal time (LST) and have low spatial resolution resulting from the wide beams of the EDGES antennas (full width at half maximum, FWHM $\geq 68^{\circ}$). The map corrections are computed by minimizing the difference between the sky measurements and the convolution of the maps with models of the EDGES antenna beams, where the free parameters are the scale and zero-level of the maps. Our map corrections complement the estimates for the spectral index of the diffuse sky in the same frequency range and from the same latitude by \citet{mozdzen2017, mozdzen2019} and \citet{mckinley2018}. To the best of our knowledge, we are the first to offer an independent correction to the G45 map, while for the LW150 map we compare our correction to suggestions by the TRIS and SARAS experiments \citep{tartari2008, patra2015}.

\section{Sky Maps}

\label{section_sky_maps}

The G45 map\footnote{\url{https://lambda.gsfc.nasa.gov/product/foreground/fg\_maipu\_info.cfm}} combines the southern sky survey by \citet{alvarez1997} at $45$~MHz with the northern sky survey by \citet{maeda1999}, which was originally done at $46.5$~MHz and then scaled to $45$~MHz. The map covers up to a declination $\delta=+65^{\circ}$. This limited sky coverage is not an impediment for our analysis since from the MRO the highest visible declination is $\delta \approx 63^{\circ}$. The authors of this map do not describe corrections for ionospheric and tropospheric effects. However, ionospheric attenuation was minimized by repeating transit observations at each declination and selecting for the map the observations affected by attenuation the least. \citet{alvarez1997} and \citet{maeda1999} report uncertainties in the map temperature scale of $10\%$ and $15\%$ respectively, while \citet{guzman2011}  suggest a zero-level correction of $-544$~K, which is $16\%$ of the lowest temperature in the map at high Galactic latitudes. In this paper we compute a correction to the original map, not the map with the zero-level correction suggested by \citet{guzman2011}.

The LW150 map\footnote{\url{https://lambda.gsfc.nasa.gov/product/foreground/fg\_all\_sky150\_mhzmap\_info.cfm}} was produced combining observations at $150$~MHz in the declination range $-25^{\circ} < \delta < +25^{\circ}$; at $85$~MHz for $\delta <-25^{\circ}$ \citep{yates1967}; and at $178$~MHz for $\delta >+25^{\circ}$ \citep{turtle1962}. Other observations were used to cover small missing regions and complete the map. The frequency scaling from $85$ and $178$~MHz to $150$~MHz was done using a single spectral index (i.e., no spatial dependence) per region. The authors of this map do not describe correcting for ionospheric and tropospheric effects. However, at these frequencies these effects are not expected to be significant compared to the overall uncertainties of the map. \citet{landecker1970} report an uncertainty of $5-7\%$ in the temperature scale and of $40$~K in the zero level, which is $28\%$ of the lowest temperature in the map at high Galactic latitudes.

\section{EDGES Data and Instruments}
\label{section_edges}

The data used to calibrate the sky maps correspond to antenna temperature measurements at $45$ and $150$~MHz conducted between 2015 and 2020 with four EDGES instruments. Details of three of the instruments are provided in \citet{monsalve2017b} and \citet{bowman2018}. Key components of the instruments include: (a) a blade dipole antenna mounted horizontally above a metal ground plane, which has a wide zenith-pointing beam; (b) an absolutely calibrated and temperature-controlled receiver mounted underneath the ground plane; (c) a back-end stage built around a $14$-bit, $400$-MS/s digitizer that yields spectra with $6.1$-kHz resolution; and (d) a vector network analyzer that conducts automated measurements of the antenna complex reflection coefficient ($S_{11}$).

\subsection{45~MHz}
\label{section_45MHz}

To calibrate the G45 map we use two datasets at $45$~MHz. The first one was obtained with the Low-Band~2 instrument described in \citet{bowman2018}. The antenna of this instrument was orientated such that the azimuth of the dipole excitation axis was $\psi_0=+87^{\circ}$. The metal ground plane consists of a central square of $20$-m side and triangular extensions of $5$-m length welded along the square perimeter, for a tip-to-tip size of $30$~m~$\times$~$30$~m. The dipole axis of the antenna was aligned parallel/perpendicular to the square outline of the ground plane. The dataset from this system corresponds to nighttime observations (with the Sun elevation below $0^{\circ}$) from days 2017-Jun-3, 2017-Oct-20, and 2018-Jan-25. These days were chosen in order to achieve the widest LST coverage possible with nighttime data. In this analysis the results are not limited by thermal noise and, therefore, once the LST coverage is maximized we do not need to integrate more days. 

The second $45$-MHz dataset was obtained in 2020 with the same Low-Band~2 system as before, except that the antenna azimuth has been changed to $\psi_0=+42^{\circ}$. The ground plane remains with the same orientation as in 2018. The change in antenna azimuth was motivated by our objective of verifying the absorption feature reported in \citet{bowman2018} with a different instrumental configuration. The rotation of the antenna relative to the fixed ground plane, which does not have rotational symmetry, produces a small but significant change in the shape of the antenna gain pattern. The rotation of the antenna relative to the local meridian impacts the spatial weighting of the sky brightness temperature by the antenna gain pattern. The data from this configuration correspond to nighttime observations from days 2020-Feb-29, 2020-Mar-29, 2020-Jul-9, and 2020-Jul-18.

\subsection{150~MHz}
\label{section_150MHz}

To calibrate the LW150 map we use two datasets at $150$~MHz. The first one was obtained with the High-Band system described in \citet{monsalve2017a, monsalve2017b} and \citet{mozdzen2017}. This instrument used a blade antenna orientated at an azimuth $\psi_0=-5^{\circ}$, and a $9.35$~m~$\times$~$9.35$~m `plus-shaped' metal ground plane aligned with the antenna. The High-Band dataset consists of nighttime observations from days 2015-Jul-25 and 2016-Jan-19.

The second dataset was obtained with the `Mid-Band' system. This instrument is a modified version of the Low-Band~1 system introduced in \citet{bowman2018}, in which the antenna has been made $\sim 25\%$ smaller (including its height above the ground plane) in order to shift its nominal frequency range to $\sim 60-150$~MHz and thus help verify the absorption feature measured with Low-Band. The azimuth of the Mid-Band antenna is $\psi_0=+85^{\circ}$ and the $30$~m~$\times$~$30$~m ground plane (which is identical to the Low-Band~2 ground plane) is aligned with the antenna. The Mid-Band dataset consists of nighttime observations from days 2018-May-26, 2018-Aug-8, and 2020-Feb-26.

\subsection{Instrument Calibration}
\label{section_calibration}

During sky observations, the input of the instruments was continuously switched between the antenna and two noise references: an attenuator acting as an ambient load, and the attenuator in series with an active noise source used to provide a higher noise level. The time spent on each of the three switch positions was $13$ seconds. We measured the power spectral density (PSD) from each position and used them to compute antenna temperature spectra in units of Kelvin at a time resolution of $39$ seconds. Finally, we brought these spectra to an absolute noise temperature scale using the formalism described in \citet{rogers2012} and \citet{monsalve2017a}. This formalism requires estimating the calibration parameters of the receiver as well as measuring the $S_{11}$ of the receiver input and of the antenna.

We determined the receiver calibration parameters from lab measurements of four external absolute calibrators connected to the receiver input in place of the antenna \citep{monsalve2017a}. Specifically, we measured the PSD, physical temperature, and $S_{11}$ of each calibrator. We then verified the receiver parameters by measuring for several hours the PSD of `antenna simulators' connected to the receiver input. These devices consist of $1.2$-m- and $2.4$-m-long cables terminated with mismatched resistive loads at ambient temperature. The mismatch and electrical length are chosen so that the simulators produce reflections comparable to those of the antennas in magnitude and phase. After applying the receiver calibration parameters and time-integrating the data, the simulator spectra are expected to be flat across frequency and to have noise temperatures equal to their time-averaged physical temperatures ($\sim 300$~K). In our verifications, the agreement between the noise temperatures and physical temperatures of the simulators was better than $100$~mK for all the receivers.

\subsection{Antenna and Ground Losses}
\label{section_losses}
After calibrating the sky observations, we removed the effect of signal loss through the balun, which is a transmission line that connects the antenna excitation port above the ground plane with the receiver input below the ground plane. We estimated the balun loss for each instrument from an analytical model of the balun, and verified it with $S_{11}$ measurements of the balun with its far end open-ended and short-circuited. 

We also corrected the observations for ground loss, which corresponds to the fraction of the antenna beam solid angle that extends below the horizon. The ground loss, as well as the antenna gain pattern above the horizon, were estimated from electromagnetic (EM) simulations with the FEKO\footnote{\url{https://www.altair.com/feko/}} (FEldberechnung bei K\"{o}rpern mit beliebiger Oberfl\"{a}che) software. The simulations included the dipole, the ground plane, and the infinite soil below the ground plane, which was modeled in terms of its conductivity and relative permittivity. 

In Table~\ref{table_antenna_parameters} we show the key parameters of the Low-Band~2, High-Band, and Mid-Band antennas.

\capstartfalse
\begin{deluxetable}{lcccc}
\tablewidth{0pt}
\tablecaption{Antenna Parameters\label{table_antenna_parameters}}
\tablehead{ & \multicolumn{2}{c}{$45$~MHz} & \multicolumn{2}{c}{$150$~MHz} \\  & Low-Band~2 & Low-Band~2 & High-Band & Mid-Band}
\startdata
Dipole Azimuth          \dotfill   &  $+42^{\circ}$                 &  $+87^{\circ}$ &  $-5^{\circ}$ &  $+85^{\circ}$ \\
Boresight Gain$^1$         \dotfill   &  $7.00$                             &  $7.00$      &  $5.89$       &  $2.88$     \\ 
Beam FWHM$^2$        \dotfill   &  $68^{\circ}\times 98^{\circ}$   &  $68^{\circ}\times 98^{\circ}$  &  $72^{\circ}\times 112^{\circ}$ &  $104^{\circ}\times 146^{\circ}$ \\
$|S_{11}|$       \dotfill   &  $-4.70$~dB                    &  $-4.78$~dB    &  $-14.19$~dB  &  $-10.96$~dB  \\
Balun Loss       \dotfill   &  $1.03\%$                      &  $1.03\%$             &  $0.88\%$            &  $0.62\%$\\
Ground Loss      \dotfill   &  $0.46\%$                      &  $0.46\%$      &  $0.48\%$     &  $0.3\%$
\enddata
\tablecomments{$^1$ Linear gain. $^2$ Parallel $\times$ perpendicular to dipole excitation axis.}
\end{deluxetable}

\subsection{Data Reduction}

For the analysis of this paper we averaged the sky observations, originally at a frequency resolution of $6.1$ kHz, into $1$-MHz bins centered at $45$ and $150$~MHz, after excising raw data affected by radio-frequency interference. We did not bin the data in time; we use them in our analysis at their original $39$-s resolution. We quantify the noise level of the frequency-binned data using an estimate for the standard deviation of the mean given by $s\cdot N_{\nu}^{-1/2}$, where $N_{\nu}$ is the number of $6.1$-kHz samples used to compute the $1$-MHz bin (typically $162$ samples) and $s$ is the sample standard deviation.

For reference, in Table~\ref{table_antenna_temperature} we report the maximum and minimum antenna temperatures measured in each of the four datasets used in this analysis. The uncertainties presented in the table correspond to $\pm 2\sigma$, which are calculated as the quadrature sum of the statistical uncertainty from noise (for a $39$-s, $1$-MHz bin) and instrumental systematic uncertainty (see Section~\ref{section_instrumental_uncertainties} for details).

\capstartfalse
\begin{deluxetable}{lcccc}
\tablewidth{0pt}
\tablecaption{Minimum and Maximum Antenna Temperatures Measured\label{table_antenna_temperature}   }
\tablehead{  & $T_{\rm A}$~[K] & LST~[h] & $T_{\rm A}$~[K] & LST~[h] \\ \\ $45$~MHz & \multicolumn{2}{c}{LB2~$+42^{\circ}$} & \multicolumn{2}{c}{LB2~$+87^{\circ}$ }}

\startdata
Minimum             &  $5607\pm 107$ & $2.56$   & $5725\pm 103$ & $2.65$  \\ 
Maximum             &  $16375\pm 308$ & $17.84$ & $17355\pm 327$ & $17.93$ \\
\hline\\
$150$~MHz & \multicolumn{2}{c}{HB~$-5^{\circ}$} & \multicolumn{2}{c}{MB~$+85^{\circ}$} \\  %
\hline
Minimum             &  $261.6\pm 0.8$ & $2.45$  & $282.1\pm 1.4$ & $1.85$ \\  
Maximum             &  $841.3\pm 4.3$ & $17.34$  & $724.3\pm 9.0$ & $17.97$
\enddata
\tablecomments{(1) LB2, HB, and MB stand for Low-Band~2, High-Band, and Mid-Band, respectively. The number next to these acronyms, such as $+42^{\circ}$, corresponds to the antenna azimuth angle. (2) Uncertainties correspond to $\pm 2\sigma$.}
\end{deluxetable}

\begin{figure*}
\begin{center}
\includegraphics[width=0.49\textwidth]{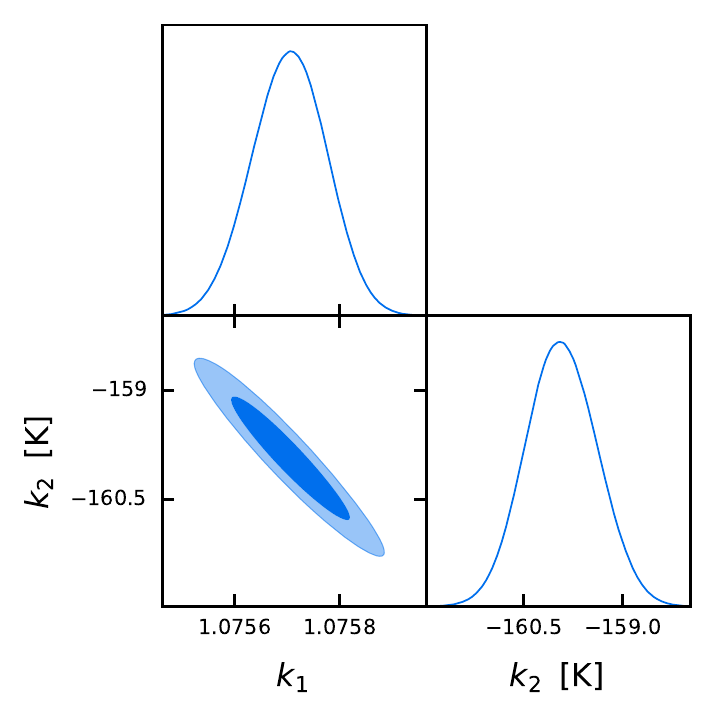}
\includegraphics[width=0.49\textwidth]{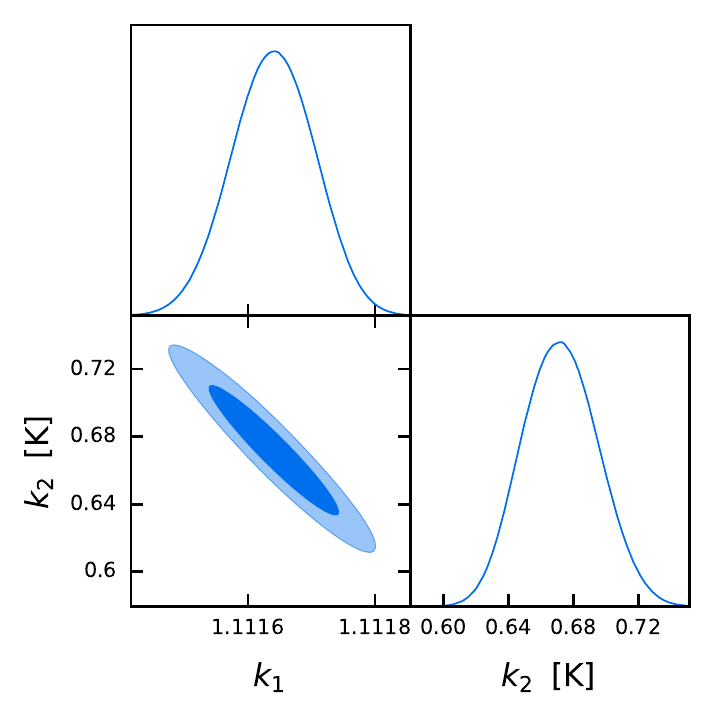}
\caption{(Left) PDFs for the correction parameters of the G45 map computed using simultaneously both $45$-MHz Low-Band 2 datasets (with antenna azimuth angles $\psi_0=+42^{\circ}$ and $+87^{\circ}$). (Right) PDFs for the correction parameters of the LW150 map computed using simultaneously both $150$-MHz datasets: High-Band with $\psi_0=-5^{\circ}$ and Mid-Band with $\psi_0=+85^{\circ}$. These PDFs encapsulate the statistical uncertainty of the estimates. We point the reader to Section~\ref{section_systematics} for a description of the systematic uncertainty.}
\label{figure_pdfs}
\end{center}
\end{figure*}

\begin{figure*}
\begin{center}
\includegraphics[width=0.99\textwidth]{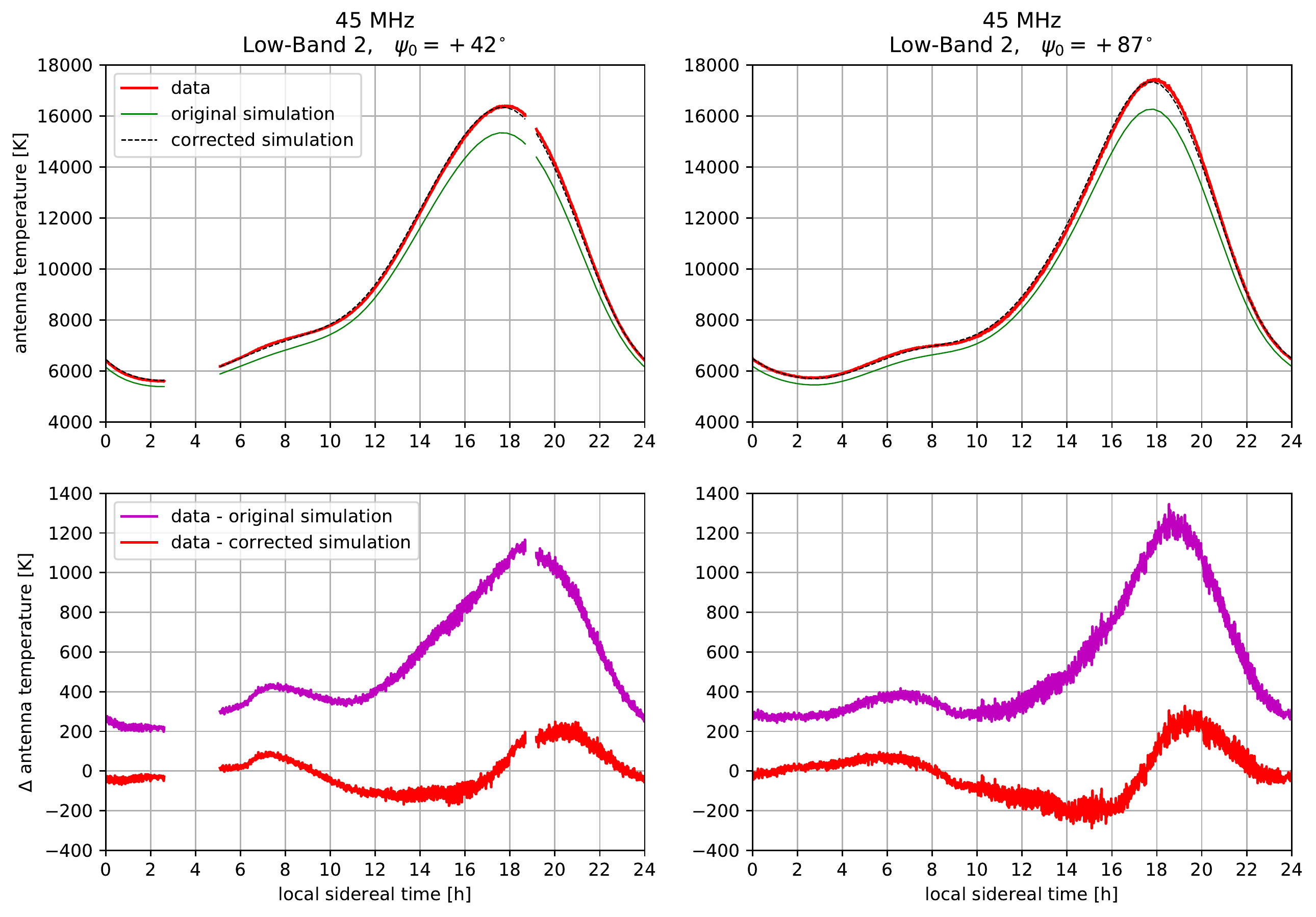}
\caption{Measurements, simulations, and residuals for the two Low-Band~2 datasets at $45$~MHz (with antenna azimuth angles $\psi_0=+42^{\circ}$ and $+87^{\circ}$). The datasets correspond to nighttime observations selected from different times of the year in order to maximize the LST coverage. (Top) The correction to the G45 map was obtained by fitting the simulated observations to both datasets simultaneously. The best-fit values are $k_1=1.076$ and $k_2=-160$~K (see Sections~\ref{section_nominal_results} and \ref{section_systematics} for descriptions of the uncertainties). The green (dashed black) lines represent the simulations without (with) the map corrections applied. (Bottom)  Difference between the data and the simulated observations. The difference decreases from the peak-to-peak range $\approx 200$-$1300$~K to within $\pm 300$~K when the map correction is applied.}
\label{figure_summary_45MHz}
\end{center}
\end{figure*}

\begin{figure*}
\begin{center}
\includegraphics[width=0.99\textwidth]{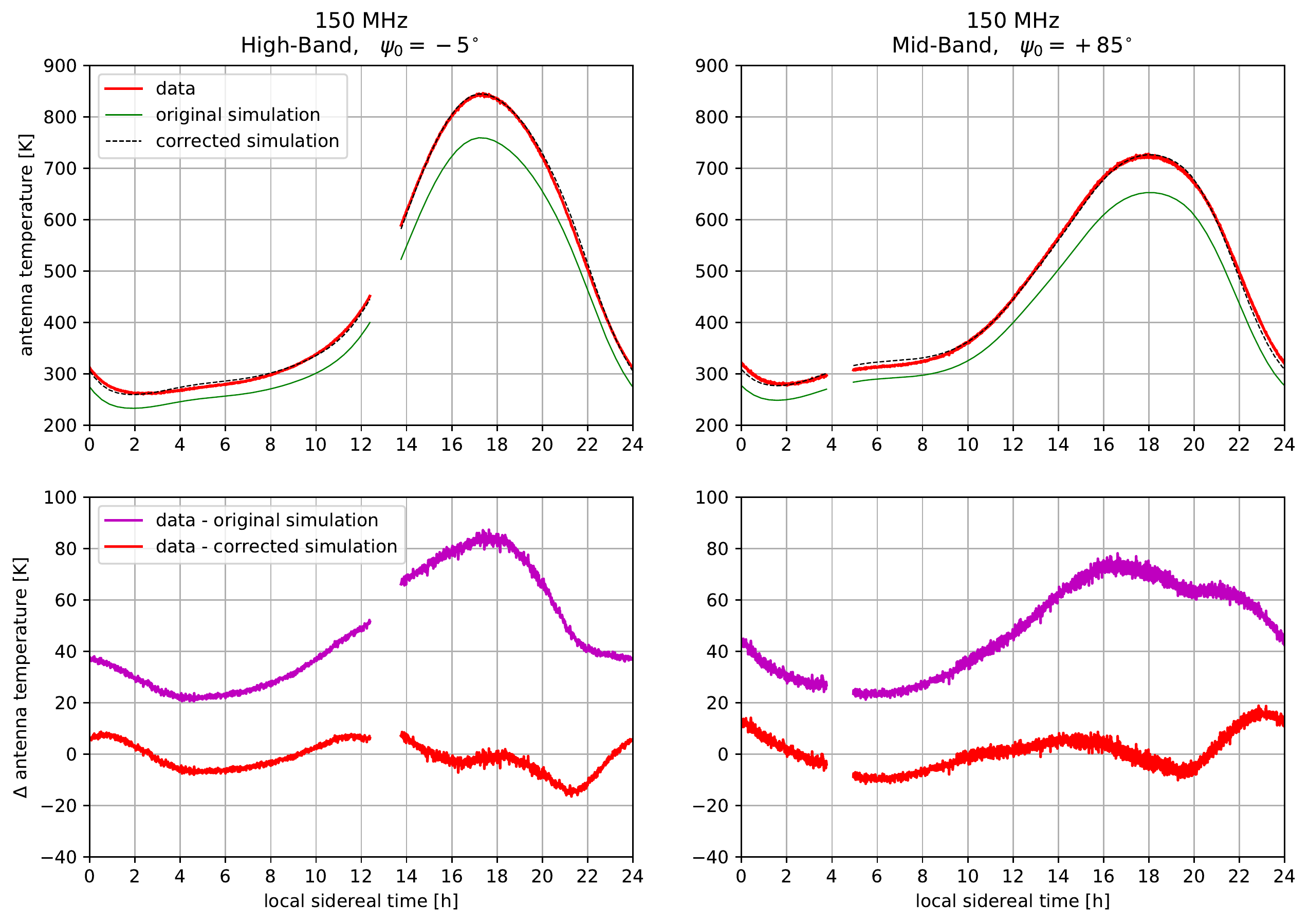}
\caption{Same as Figure~\ref{figure_summary_45MHz} but for $150$~MHz. The datasets are from High-Band with antenna azimuth $\psi_0=-5^{\circ}$ and Mid-Band with $\psi_0=+85^{\circ}$. (Top) The correction to the LW150 map was obtained by fitting the simulated observations to both datasets simultaneously. The best-fit values are $k_1=1.112$ and $k_2=0.7$~K. The green (dashed black) lines represent the simulations without (with) the map corrections applied. (Bottom) Difference between the data and the simulated observations. The difference decreases from the peak-to-peak range $\approx 20$-$90$~K to within $\pm 20$~K when the map correction is applied.}
\label{figure_summary_150MHz}
\end{center}
\end{figure*}

\section{Simulated Observations}
\label{section_simulations}

The calibrated antenna temperature measurements described in Section~\ref{section_edges} correspond to the convolution of the sky brightness temperature with the antenna gain pattern. Following the same prescription, we compute simulated antenna temperature measurements, $\hat{T}_{\rm A}$, as:

\begin{align}
& \hat{T}_{\rm A}({\rm LST}) = \nonumber \\
& \frac{1}{4\pi} \int_{\phi=0}^{2\pi}\int_{\theta=0}^{\frac{\pi}{2}} G(\theta,\phi; \psi_0)\cdot T_{\rm sky}(\theta,\phi; {\rm LST}) \cdot \sin\theta d\theta d\phi,
\label{equation_convolution}
\end{align}

\noindent where $G$ is our model for the normalized antenna gain pattern, $\phi$ and $\theta$ are the azimuth and zenith angles, respectively, $\psi_0$ is the azimuth angle of the dipole excitation axis, and $T_{\rm sky}$ is the diffuse radio map to be corrected. Specifically, we compute simulated antenna temperatures at $45$~MHz by convolving the two Low-Band~2 gain pattern models from EM simulations rotated in azimuth according to Table~\ref{table_antenna_parameters}, with the G45 map, and at $150$~MHz by convolving the azimuth-rotated High-Band and Mid-Band gain pattern models with the LW150 map. The simulations are done for an observation latitude of $-26.7^{\circ}$ (MRO). We compute simulated antenna temperatures with Equation~\ref{equation_convolution} every $20$~minutes across $24$~hours of LST, and then use interpolation in order to obtain values at the LSTs of the actual sky measurements.

\section{Results}
\label{section_results}

\subsection{Parameter Fits}
\label{section_parameter_fits}

We assume the following model for the correction of the G45 and LW150 maps:

\begin{equation}
\hat{T}_{\rm A}^* = k_1\cdot\hat{T}_{\rm A} + k_2,
\label{equation_correction}
\end{equation}

\noindent where $\hat{T}_{\rm A}$ is the simulated antenna temperature from Section~\ref{section_simulations}, $\hat{T}_{\rm A}^*$ is the corrected simulation, and $k_1$ and $k_2$ are the temperature scale and zero-level corrections respectively. To fit the parameters we use the PolyChord implementation of the Nested Sampling algorithm \citep{handley2015a, handley2015b}. Our log-likelihood function is

\begin{equation}
\ln \mathscr{L}  = -\frac{1}{2}\cdot \left[T_{\rm A}-\hat{T}_{\rm A}^*(k_1, k_2)\right]^T \Sigma^{-1} \left[T_{\rm A}-\hat{T}_{\rm A}^*(k_1, k_2)\right],
\label{equation_likelihood}
\end{equation}

\noindent where $T_{\rm A}$ represents the measurements and $\Sigma$ is the diagonal noise covariance matrix of size $N_{\rm LST}\times N_{\rm LST}$, where $N_{\rm LST}$ is the number of points across LST. We adopt uniform parameter priors that span the ranges $[10^{-2}, 10^{+2}]$ for $k_1$ and $[-10^5, +10^5]$~K for $k_2$.

\subsection{Nominal Results}
\label{section_nominal_results}

Our nominal results for the G45 map are derived from the simultaneous fit of $k_1$ and $k_2$ to the two datasets at $45$~MHz; i.e., in Equation~\ref{equation_likelihood} $T_{\rm A}$ represents the concatenation of both datasets and, correspondingly, $\hat{T}^*_{\rm A}$ is the concatenation of both simulations. The results are shown in the left triangle plot of Figure~\ref{figure_pdfs} in the form of 1D and 2D probability density functions (PDFs) for $k_1$ and $k_2$. The best-fit values are $k_1=1.07570\pm 0.00007$ and $k_2=-159.9\pm 0.5$~K, where the limits enclose the $68\%$ confidence ranges due to statistical uncertainty.

Our nominal results for the LW150 map come from the simultaneous fit to the two datasets at $150$~MHz from High-Band and Mid-Band. The PDFs are shown in the right triangle plot of Figure~\ref{figure_pdfs} and the best-fit values and $68\%$ ranges are $k_1=1.11164\pm 0.00006$ and $k_2=0.67\pm 0.02$~K.

Figure~\ref{figure_summary_45MHz} shows the data, simulations, and residuals for the nominal analysis at $45$~MHz. The top panels show the measured antenna temperature, the simulations with the original G45 map, and the simulations with the corrected map after applying the best-fit $k_1$ and $k_2$. The bottom panels show the difference between the data and the original and corrected simulations. Here we can see the significant reduction of the residuals, from the original range $\approx 200-1300$~K to within $\pm 300$~K when the correction is applied. The root-mean-square (RMS) of the residuals after correction is $101$~K ($115$~K) for Low-Band~2 with antenna azimuth $\psi_0=+42^{\circ}$ ($+87^{\circ}$). Figure~\ref{figure_summary_150MHz} shows the data, simulations, and residuals for the nominal analysis at $150$~MHz. Here, the residuals go from the range $\approx 20-90$~K when using the original maps, to within $\pm 20$~K after correction. The RMS of the residuals after correction is $5.9$~K ($6.5$~K) for the High-Band (Mid-Band) data.

\subsection{Systematic Uncertainties}
\label{section_systematics}

Until this point we have not accounted for systematic uncertainty in the estimates for $k_1$ and $k_2$, which could arise due to uncertainties in instrument modeling or calibration, ionospheric and tropospheric effects, or our choice of analysis strategy. We address these effects in this subsection. 

We estimate the total systematic uncertainty, $\Delta$, on each of the two parameters using the following quadrature sum:

\begin{align}
\Delta^2 &= \Delta^2_{\rm dataset} +\Delta^2_{\rm refr} + \Delta^2_{\rm iae}\nonumber \\
&+\Delta^2_{\rm recv,D1} + \Delta^2_{\rm S_{11},D1} + \Delta^2_{\rm balun,D1} + \Delta^2_{\rm bgl,D1} \nonumber \\
&+ \Delta^2_{\rm \psi_0,D1} + \Delta^2_{\rm ra,D1} + \Delta^2_{\rm dec,D1} \nonumber \\
&+\Delta^2_{\rm recv,D2} + \Delta^2_{\rm S_{11},D2} + \Delta^2_{\rm balun,D2} + \Delta^2_{\rm bgl,D2} \nonumber \\
&+ \Delta^2_{\rm \psi_0,D2} + \Delta^2_{\rm ra,D2} + \Delta^2_{\rm dec,D2}. \nonumber \\
\label{equation_systematic_uncertainty}
\end{align}

Here, each term represents uncertainty in the estimated parameter ($k_1$ or $k_2$) due to a different uncertainty source. This is an approximation to the true systematic uncertainty under the realistic assumption that the source uncertainties are uncorrelated. Note that, in general, the various uncertainty sources would contribute in a non-linear manner, rendering this approximation a first-order estimate only, especially for large values of the terms. Furthermore, many of the terms here do not have well-understood statistical distributions, and we shall be forced to make judgments about their variance from limited data. In this, we shall attempt to be conservative. A more rigorous analysis would forward-model all effects simultaneously from prior distributions that have been sufficiently validated theoretically or empirically, but we postpone such an analysis to future work.

Below we describe the terms contributing to Equation~\ref{equation_systematic_uncertainty} and the source uncertainty assumed for them. We choose to report uncertainties at a level regarded as $2\sigma$; although not rigorous in a statistical sense, the systematic uncertainty ranges we report are expected to contain the true value with high probability. In Table~\ref{table_systematics} we show the values of the uncertainty terms in Equation~\ref{equation_systematic_uncertainty} for the correction parameters of both maps.

\capstartfalse
\begin{deluxetable}{lcccc}
\tablewidth{0pt}
\tablecaption{Systematic Uncertainties in the Correction Parameters of the G45 and LW150 Maps \label{table_systematics}}
\tablehead{ G45 & \multicolumn{2}{c}{$k_1$} & \multicolumn{2}{c}{$k_2$~[K]} }
\startdata                  &  & \\
$\Delta_{\rm dataset}$  \dotfill   &  \multicolumn{2}{c}{0.0026} & \multicolumn{2}{c}{43.3} \\
$\Delta_{\rm refr}$     \dotfill   &  \multicolumn{2}{c}{0.0013} & \multicolumn{2}{c}{14.7} \\
$\Delta_{\rm iae}$      \dotfill   &  \multicolumn{2}{c}{0.0164} & \multicolumn{2}{c}{3.9} \\
\\
 & LB2~$+42^{\circ}$ & LB2~$+87^{\circ}$ & LB2~$+42^{\circ}$ & LB2~$+87^{\circ}$\\
$\Delta_{\rm recv}$    \dotfill   & 0.0205              & 0.0205  &  5.0  &  10.1 \\
$\Delta_{\rm S_{11}}$  \dotfill   & 0.0003              & 0.0002  &  5.4  &   0.1 \\
$\Delta_{\rm balun}$   \dotfill   & 0.0011              & 0.0011  &  6.0  &   0.5 \\
$\Delta_{\rm bgl}$     \dotfill   & 0.0004              & 0.0004  &  0.9  &   0.7 \\
$\Delta_{\rm \psi_0}$      \dotfill   & 0.0008              & 0.0008  &  3.2  &   7.8 \\
$\Delta_{\rm ra}$      \dotfill   & 0.0039              & 0.0067  & 42.5  &  43.8 \\
$\Delta_{\rm dec}$     \dotfill   & 0.0038              & 0.0019  &  5.8  &   2.1 \\            
\\
$\Delta$               &  \multicolumn{2}{c}{0.0341} & \multicolumn{2}{c}{78.3} \\
\\
\hline
\hline
\\
LW150  & \multicolumn{2}{c}{$k_1$} & \multicolumn{2}{c}{$k_2$~[K]} \\ 
\hline                     &  &  \\
$\Delta_{\rm dataset}$  \dotfill   &  \multicolumn{2}{c}{0.0087} & \multicolumn{2}{c}{0.48} \\
$\Delta_{\rm refr}$     \dotfill   &  \multicolumn{2}{c}{0.0014} & \multicolumn{2}{c}{0.50} \\
$\Delta_{\rm iae}$      \dotfill   &  \multicolumn{2}{c}{0.0014} & \multicolumn{2}{c}{1.32} \\
\\
 & HB~$-5^{\circ}$ & MB~$+85^{\circ}$ & HB~$-5^{\circ}$ & MB~$+85^{\circ}$\\
$\Delta_{\rm recv}$     \dotfill   & 0.0082                  & 0.0159  &  2.36 &  3.72  \\
$\Delta_{\rm S_{11}}$   \dotfill   & 0.0001                  & 0.0000  &  0.00 &  0.01  \\
$\Delta_{\rm balun}$    \dotfill   & 0.0010                  & 0.0007  &  0.26 &  0.19  \\
$\Delta_{\rm bgl}$      \dotfill   & 0.0001                  & 0.0047  &  0.24 &  1.63  \\
$\Delta_{\rm \psi_0}$       \dotfill   & 0.0018                  & 0.0037  &  0.79 &  1.64  \\
$\Delta_{\rm ra}$       \dotfill   & 0.0038                  & 0.0012  &  0.84 &  0.43  \\ 
$\Delta_{\rm dec}$      \dotfill   & 0.0041                  & 0.0059  &  0.42 &  2.67  \\ 
\\
$\Delta$             &  \multicolumn{2}{c}{0.0228} & \multicolumn{2}{c}{6.00}
\enddata
\tablecomments{(1) We point the reader to Section~\ref{section_systematics} for a description of the systematic uncertainties. (2) G45 refers to the \citet{guzman2011} $45$-MHz map. LW150 refers to the \citet{landecker1970} $150$-MHz map. (3) LB2, HB, and MB stand for Low-Band~2, High-Band, and Mid-Band, respectively. The number next to these acronyms, such as $+42^{\circ}$, corresponds to the antenna azimuth angle. (4) The total systematic uncertainty in the parameter estimates is denoted as $\Delta$ and is computed following Equation~\ref{equation_systematic_uncertainty}. (5) In Equation~\ref{equation_systematic_uncertainty}, D1 and D2 correspond to the two datasets used at each frequency. At $45$~MHz, the two datasets are LB2~$+42^{\circ}$ and LB2~$+87^{\circ}$. At $150$~MHz, the two datasets are HB~$-5^{\circ}$ and MB~$+85^{\circ}$.}
\end{deluxetable}

\subsubsection{Choice of Datasets}
\label{section_instrumental_dataset}

The first term in Equation~\ref{equation_systematic_uncertainty}, $\Delta_{\rm dataset}$, represents uncertainty in the parameter estimates due to our choice for the dataset used in the analysis. We compute this term as the absolute difference between the parameter values obtained when we fit them separately to each of the two datasets (D1 and D2) considered for each map, instead of simultaneously as for the nominal results. This uncertainty term tests for potential inconsistencies between the two datasets due to otherwise unaccounted effects, such as calibration errors not captured by the other terms. This term is also sensitive to potential variations of the scale and zero-level across coordinates in the maps, which become apparent when they are convolved with a beam rotated to a different azimuth angle.

\subsubsection{Ionospheric and Tropospheric Effects}
\label{section_ionosphere}

The Earth's ionosphere impacts incoming radio waves through attenuation, thermal emission, and refraction \citep{vedantham2014, rogers2015, sokolowski2015, datta2016, shen2020}. Further, at low altitude radio waves are also refracted by the troposphere \citep{vedantham2014}. In Equation~\ref{equation_systematic_uncertainty}, the term $\Delta_{\rm refr}$ represents uncertainty due to refraction by the troposphere and ionosphere, and $\Delta_{\rm iae}$ is the uncertainty due to attenuation and emission by the ionosphere.

As mentioned in Section~\ref{section_sky_maps}, the authors of the G45 and LW150 maps do not describe having made corrections for tropospheric and ionospheric effects. For the G45 map, they only describe choosing observations with low ionospheric attenuation. Because of this, in this paper we do not remove tropospheric and ionospheric effects from our observations before calibrating the maps. Nonetheless, the data we use here correspond to nighttime observations and, thus, the impact of the ionosphere is expected to be low. 

Although both the maps and our observations are affected by tropospheric and ionospheric effects, most likely the levels of attenuation, emission, and refraction affecting them are different. The uncertainty terms $\Delta_{\rm refr}$ and $\Delta_{\rm iae}$ account for these differences. They are computed as the absolute difference in the estimates for $k_1$ and $k_2$ between the nominal values and results obtained after including the tropospheric and ionospheric effects in the simulated observations. In the nominal simulations we implicitly assume that the maps are already affected by these effects and that the effects are equal to those affecting our data. In the alternative simulations, on the other hand, we explicitly incorporate the tropospheric and ionospheric effects under the assumption that the G45 and LW150 maps represent the sky temperature as seen from outside the ionosphere. Details of the alternative simulations are prrovided in the Appendix.

We assume the same tropospheric and ionospheric conditions in all the alternative simulations and, in particular, typical nighttime values for the ionospheric parameters. Since the assumed conditions are the same across simulations, we estimate the alternative correction parameters for each map by fitting the alternative simulations to both datasets simultaneously, in the same way as for the nominal simulations. Similarly to the other effects, $\Delta_{\rm refr}$ and $\Delta_{\rm iae}$ computed as just described represent uncertainties at the $2\sigma$ level.

Ionospheric effects are stronger at lower frequencies and hence, as Table~\ref{table_systematics} shows, they have a larger impact on the corrections of the G45 map. We see that $k_1$ is impacted more significantly by ionospheric attenuation and emission, while $k_2$ is impacted more significantly by tropospheric and ionospheric refraction.

\subsubsection{Instrumental Uncertainties}
\label{section_instrumental_uncertainties}

The next fourteen terms in Equation~\ref{equation_systematic_uncertainty} represent instrumental uncertainties affecting independently the two datasets. We take this `per dataset' approach because in general these uncertainties affect the observations from each instrument differently. These uncertainties are measured with respect to the results for the same dataset with nominal calibration.

The term $\Delta_{\rm recv}$ (for D1 and D2) represents uncertainty in the parameter estimates due to uncertainty in receiver calibration. We compute this term as the absolute difference between the nominal estimates and the results obtained when we calibrate the data using receiver parameters derived from a different set of lab measurements (spectra, reflection coefficients, and physical temperatures of the absolute calibrators). For Low-Band~2, the two receiver calibrations considered were done in 2018-Sep and 2019-Dec (nominal); for High-Band the two calibrations are from 2015-Mar and 2017-Jan (nominal); and for Mid-Band, the two calibrations are from 2018-Jan (nominal) and 2019-Nov.  

The term $\Delta_{\rm S_{11}}$ accounts for uncertainty in the antenna $S_{11}$. We compute it as the absolute difference between the nominal parameter estimates and the results obtained when we calibrate the data using an $S_{11}$ affected by a realistic error. We obtain this $S_{11}$ by perturbing the nominal $S_{11}$ measured in the field; specifically, we add the complex value $2(1+i)\times 10^{-4}$, which represents a realistic error in our calibrated $S_{11}$ measurements at the $2\sigma$ level \citep{monsalve2017a}.

The term $\Delta_{\rm balun}$ accounts for uncertainty in the balun loss. We compute it as the absolute difference between the nominal estimates and the values obtained when we apply a perturbation to the nominal loss, which is determined from theoretical models and verified through measurements. Conservatively, to represent uncertainty at the $2\sigma$ level we apply a perturbation of $10\%$ of the nominal value.

The term $\Delta_{\rm bgl}$ accounts for uncertainty in our antenna beam pattern model as well as in our ground loss estimate. The beam pattern is primarily determined by the physical characteristics of the antenna but is also very sensitive to the properties of the metal ground plane and soil. Further, the beam solid angle above the horizon is complementary to the ground loss fraction since they add up to $4\pi$. To account for potential imperfections in our beam model and ground loss estimate, we apply perturbations to our simulated observations. Specifically, to represent uncertainty at the $2\sigma$ level, the perturbations correspond to $10\%$ of the difference between the nominal simulations and an idealized case in which the antenna model has an infinite metal ground plane with zero resistivity. This antenna model produces a different beam pattern above the horizon compared to the nominal case, as well as zero ground loss. The term $\Delta_{\rm bgl}$, then, is computed as the absolute difference between the nominal estimates for $k_1$ or $k_2$ and the estimates obtained for the perturbed simulations.

The term $\Delta_{\rm \psi_0}$ accounts for uncertainty in the antenna azimuth angle. We compute it as the absolute difference between the nominal parameter estimates and the estimates obtained when we assume a $2\sigma$ azimuth error of $1^{\circ}$. This error is introduced in the simulated observations as a perturbation to the nominal antenna azimuth. 

The terms $\Delta_{\rm ra}$ and $\Delta_{\rm dec}$ account for uncertainty in the antenna pointing, which nominally corresponds to the zenith. `ra' and `dec' stand for right ascension and declination, and we assume $2\sigma$ uncertainties of $1^{\circ}$ for both. This is supported by field measurements of the antenna panels and metal ground planes, which indicate that they depart from perfectly level by $<0.5^{\circ}$. Specifically, $\Delta_{\rm ra}$ is computed as the absolute difference between the nominal map correction parameters and those obtained when using a simulation in which the LST is shifted by $1^{\circ}$~($4$~min) relative to the nominal. To compute $\Delta_{\rm dec}$, in the simulation we assume an observation latitude shifted by $1^{\circ}$ relative to the nominal.

\subsubsection{Sky Polarization Effects}
\label{section_polarization}

Using the diffuse polarization simulations from \citet{spinelli2018}, \citet{spinelli2019} estimated the contribution from sky polarization to measurements with wide-beam single-polarization dipole antennas below $200$~MHz. They found that for short integrations in LST this contribution is within $\pm 1$~K for all LSTs\footnote{Private communication with Marta Spinelli.}. This effect is thus smaller than the noise level of the data used in this analysis and is negligible when compared to the differences between our data and the corrected simulations. As a result, sky polarization has a negligible impact on our coordinate-independent corrections to the maps when compared to the effects previously discussed. For this reason we do not include a term to account for sky polarization in Equation~\ref{equation_systematic_uncertainty}.

\subsubsection{Total Uncertainties}

After adding in quadrature all the systematic uncertainty terms in Equation~\ref{equation_systematic_uncertainty} we obtain the total value, $\Delta$, which is shown in the last row of the G45 and LW150 sections of Table~\ref{table_systematics}. We then calculate the total uncertainties by combining in quadrature the total systematic uncertainties and the statistical uncertainties found in our nominal analysis (Section~\ref{section_nominal_results}). Because the statistical uncertainties are smaller by orders of magnitude, the systematic uncertainties dominate the total values. Our final results for the temperature scale and zero-level, including the total uncertainties ($2\sigma$), are $k_1=1.076\pm0.034$ and $k_2=-160\pm 78$~K for the G45 map and $k_1=1.112\pm 0.023$ and $k_2=0.7\pm 6.0$~K for the LW150 map.

\begin{figure*}
\begin{center}
\includegraphics[width=0.99\textwidth]{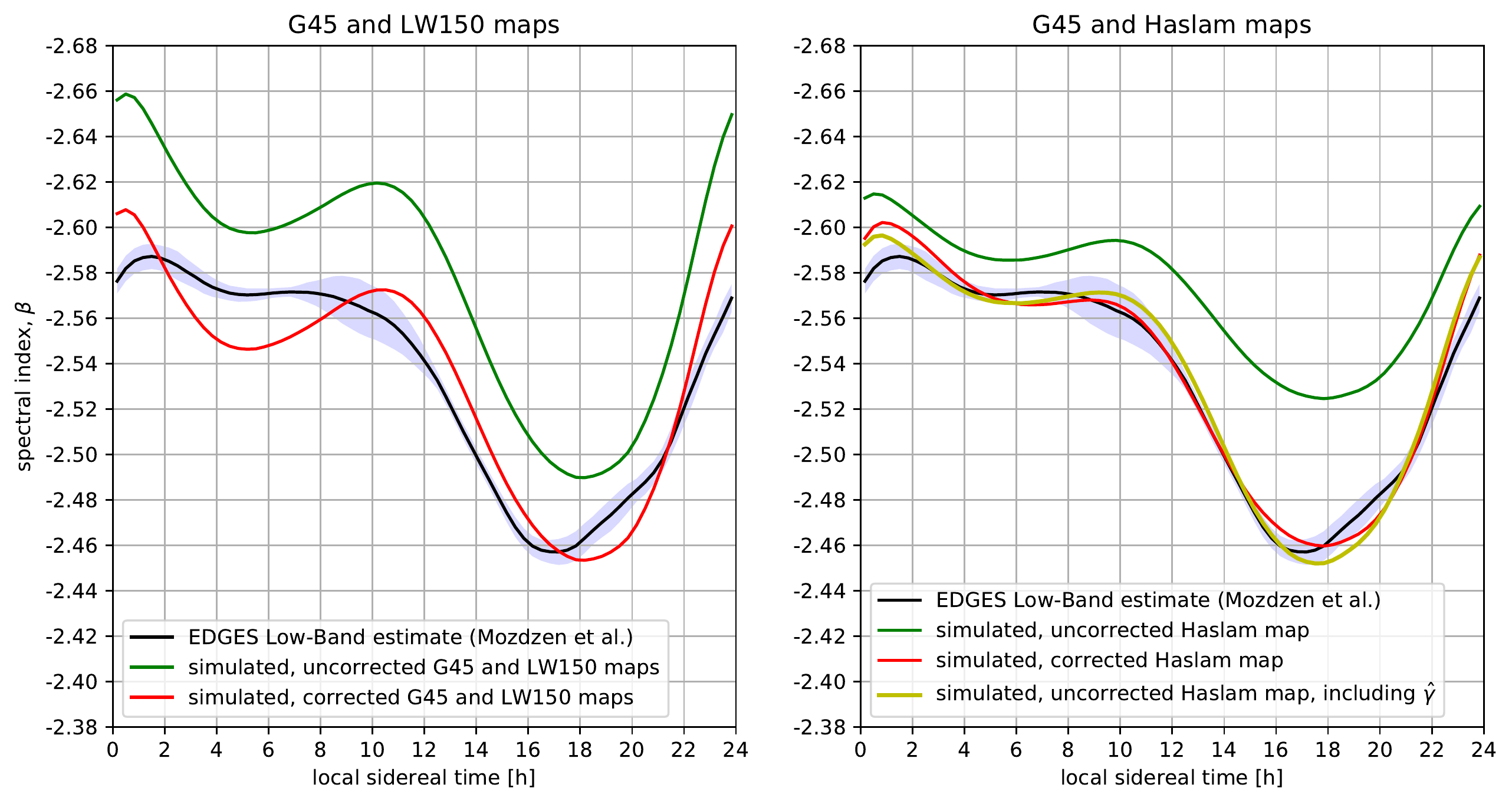}
\caption{(Left) Comparison between the spectral index of the diffuse radio sky estimated from EDGES data in the range $50-100$~MHz \citep{mozdzen2019}, with simulations computed with Equation~\ref{equation_spectral_index}. The solid black line represents the data. The light blue region represents the $2\sigma$ uncertainty band of the data. Comparing the green and red lines we can see that the simulations agree better with the data when the corrections are applied to the G45 and LW150 maps. (Right) Same as the left panel but computing the simulated spectral index using the corrected G45 map and the $408$-MHz Haslam map \citep{haslam1981,haslam1982,remazeilles2015}. We see that a very good agreement with the data can be obtained --- better than when using the LW150 map --- if we apply corrections to the Haslam map or when we incorporate into the simulation an LST-dependent curvature ($\hat{\gamma}$). Due to the very strong degeneracy between the two alternatives and the lack of precise constraints on the curvature in the $45-408$~MHz range, we cannot determine robust corrections to the Haslam map from our lower-frequency data.}
\label{figure_consistency_check}
\end{center}
\end{figure*}

\section{Consistency Check}

\subsection{Spectral Index for $45-150$~MHz}

To crosscheck the accuracy of our map corrections, we compare the spectral index, $\hat{\beta}$, derived from simulated observations of the corrected G45 and LW150 maps, with the $\beta$ estimated in \citet{mozdzen2019} from Low-Band data in the range $50-100$~MHz. The Low-Band~2 observations used here to correct the G45 map come from the same dataset used in \citet{mozdzen2019}; therefore, this crosscheck is not completely independent. However, here we use data only at $45$~MHz instead of at $50-100$~MHz. Furthermore, our correction for the LW150 map relies on independent data from the High- and Mid-Band systems, which makes this crosscheck at least semi-independent and, thus, valuable.

The nominal $\beta$ reported in \citet{mozdzen2019} represents the variation with frequency of the sky brightness temperature without removing nighttime ionospheric and tropospheric effects and after removing the contribution from the CMB. Therefore, to make the comparison valid, here we compute $\hat{\beta}$ as

\begin{equation}
\hat{\beta} = \frac{\ln\left(\frac{\hat{T}^*_{{\rm A},150\rm MHz} - T_{\rm CMB}}{\hat{T}^*_{{\rm A},45\rm MHz} - T_{\rm CMB}}\right)}{ \ln\left(\frac{150\; \rm MHz}{45\; \rm MHz}\right)},
\label{equation_spectral_index}
\end{equation}

\noindent where $T_{\rm CMB}=2.725$~K \citep{mather1999}, and $\hat{T}^*_{A,45\rm MHz}$ and $\hat{T}^*_{A,150\rm MHz}$ are the simulated antenna temperatures computed using Equation~\ref{equation_convolution} and the corrected maps. Two more aspects of this comparison are worth mentioning. First, the $\beta$ reported in \citet{mozdzen2019} was obtained after applying a correction to the data that removed the effect of variations of the beam gain with frequency relative to a reference value of $75$~MHz. Therefore, in this check we compute the simulated antenna temperatures at $45$ and $150$~MHz using the EDGES Low-Band beam gain model at $75$~MHz ($G$ at $75$~MHz in Equation~\ref{equation_convolution}). Second, in \citet{mozdzen2019} the reported $\beta$ is the average of results from observations with antenna azimuth angles of $-7^{\circ}$ and $+87^{\circ}$. To reproduce this aspect here, the final $\hat{\beta}$ is obtained by averaging the values from simulations at those two angles. 

The results of this check are presented in the left panel of Figure~\ref{figure_consistency_check}. We can see that the agreement of the simulation with the data improves significantly (considering the data uncertainties) when the corrections to the maps are applied. Upon correction, $\hat{\beta}$ is less steep (less negative values) by $\approx 0.03-0.05$ across LST, which occurs because our scale correction for the G45 map ($+7.6\%$) is lower than that for the LW150 map ($+11.2\%$).

\subsection{Spectral Index for $45-408$~MHz}

Although the agreement between the data and the simulation improves when correcting the maps, in the left panel of Figure~\ref{figure_consistency_check} we see that there are still significant disagreements between them, especially at LST$\sim 22$ to $10$~h. To explore if these differences could be produced by coordinate-dependent errors in the maps, we replace one of the maps with the $408$-MHz Haslam map\footnote{Specifically, we use the destriped and not desourced \citet{remazeilles2015} version of the Haslam map. The differences seen when using other versions of the map are negligible.} \citep{haslam1981, haslam1982, remazeilles2015}, which is the only other radio map with full sky coverage. We first compute $\hat{\beta}$ using the Haslam and LW150 maps (not using the G45 map), but the differences with the data at LST$\sim 22$ to $10$~h remain significant. We then compute $\hat{\beta}$ using the Haslam and G45 maps (not using the LW150 map) and the agreement improves. We show this case in the right panel of Figure~\ref{figure_consistency_check} using the green line. Even though there is an offset of $\approx 0.01-0.04$ between the simulation and the data at LST$\sim 22$ to $10$~h, and larger at the other LSTs, the variations with LST are similar.

In the previous case we used the original Haslam map without any modification. To see if we could improve the agreement with the data, we then applied corrections to the Haslam map. We determined the corrections by directly fitting $\hat{\beta}$ (which is a function of the scale and zero-level of the Haslam map) to $\beta$ from \citet{mozdzen2019} considering its uncertainties. We found best-fit corrections of $+1.21$ to the scale and of $-4.1$~K to the zero-level of the Haslam map. We show $\hat{\beta}$ after applying this correction in the right panel of Figure~\ref{figure_consistency_check} using the red line. We can see that this $\hat{\beta}$ is a significant improvement relative to the previous cases. The discrepancies between the simulation and the data have been reduced to the same magnitude as the uncertainty in the data (the light blue regions represent $2\sigma$ uncertainties). Although the agreement is good, we caution that this result was obtained under the extreme assumption that the spectral dependence between $45$ and $408$~MHz can be perfectly modeled using only a spectral index (albeit with coordinate dependence).

We then tried to improve the agreement relative to the case with the original Haslam map by, instead of modifying the Haslam map, assuming an appropriate non-zero value for the `curvature' of the spectral index, $\gamma$. Motivated by the middle panel of Figure~$7$ in \citet{mozdzen2019}, we used a template for the dependence of $\gamma$ with LST that is a scaled version of the antenna temperature itself. We determined the scaling parameters by fitting to the data the following equation for $\hat{\beta}$:

\begin{equation}
{\hat \beta} = \frac{\ln\left(\frac{\hat{T}_{{\rm A},408\rm MHz} - T_{\rm CMB}}{\hat{T}^*_{{\rm A},45\rm MHz} - T_{\rm CMB}}\right)  - \hat{\gamma} \left[\ln\left(\frac{408\; \rm MHz}{45\; \rm MHz}\right)\right]^2}{ \ln\left(\frac{408\; \rm MHz}{45\; \rm MHz}\right)}.
\label{equation_spectral_index_with_gamma}
\end{equation}

\begin{figure*}
\begin{center}
\includegraphics[width=0.49\textwidth]{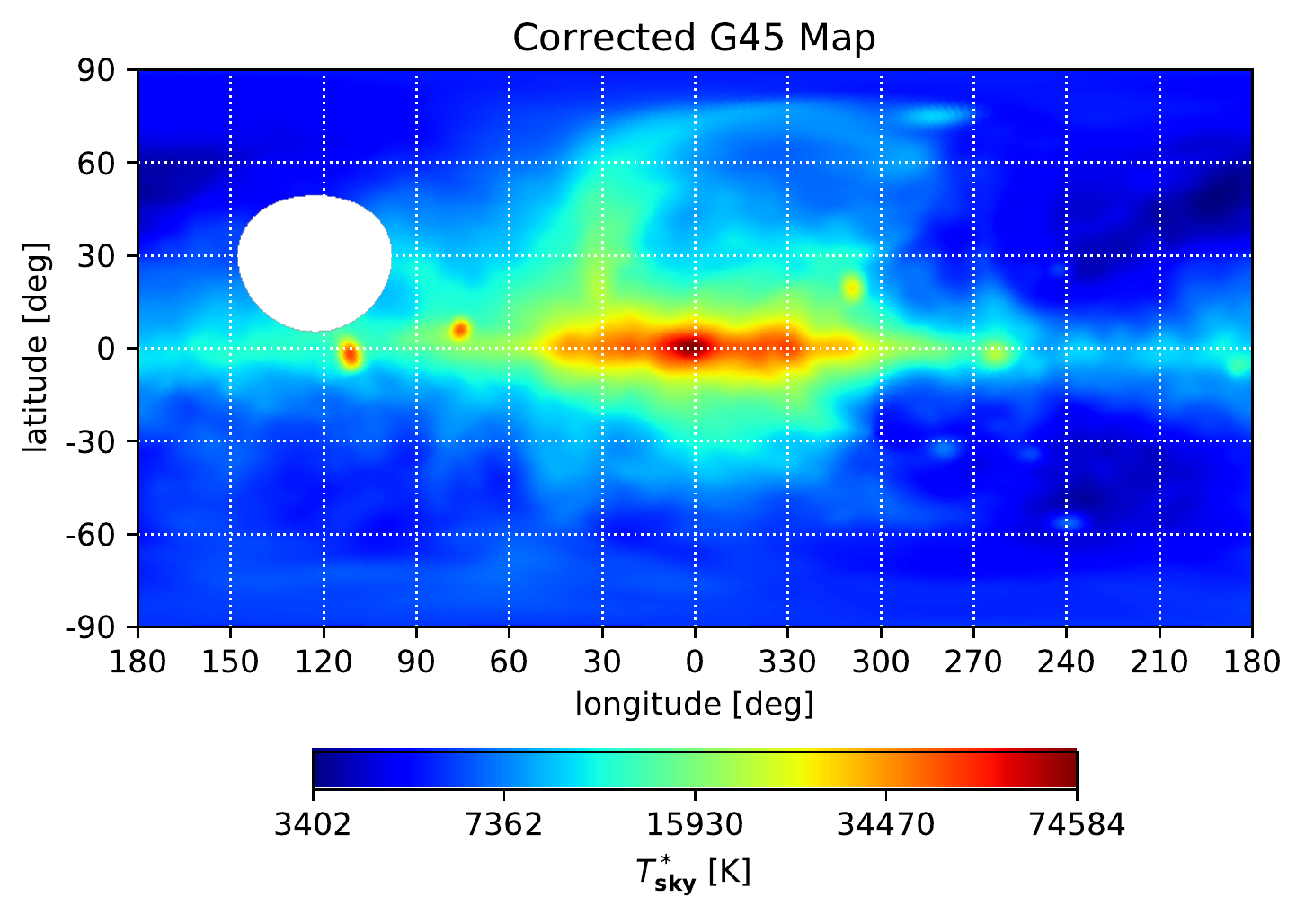}
\includegraphics[width=0.49\textwidth]{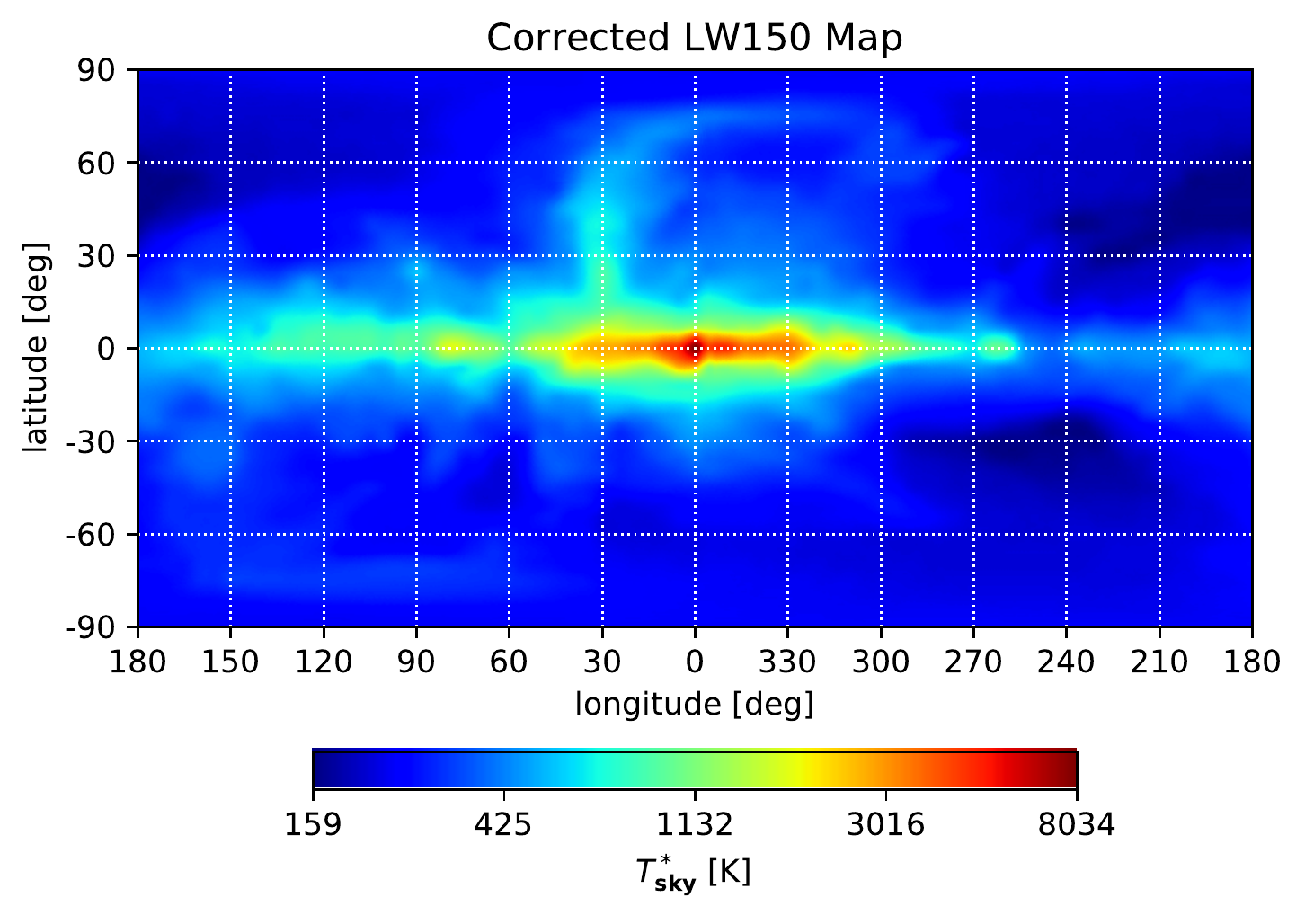}
\includegraphics[width=0.49\textwidth]{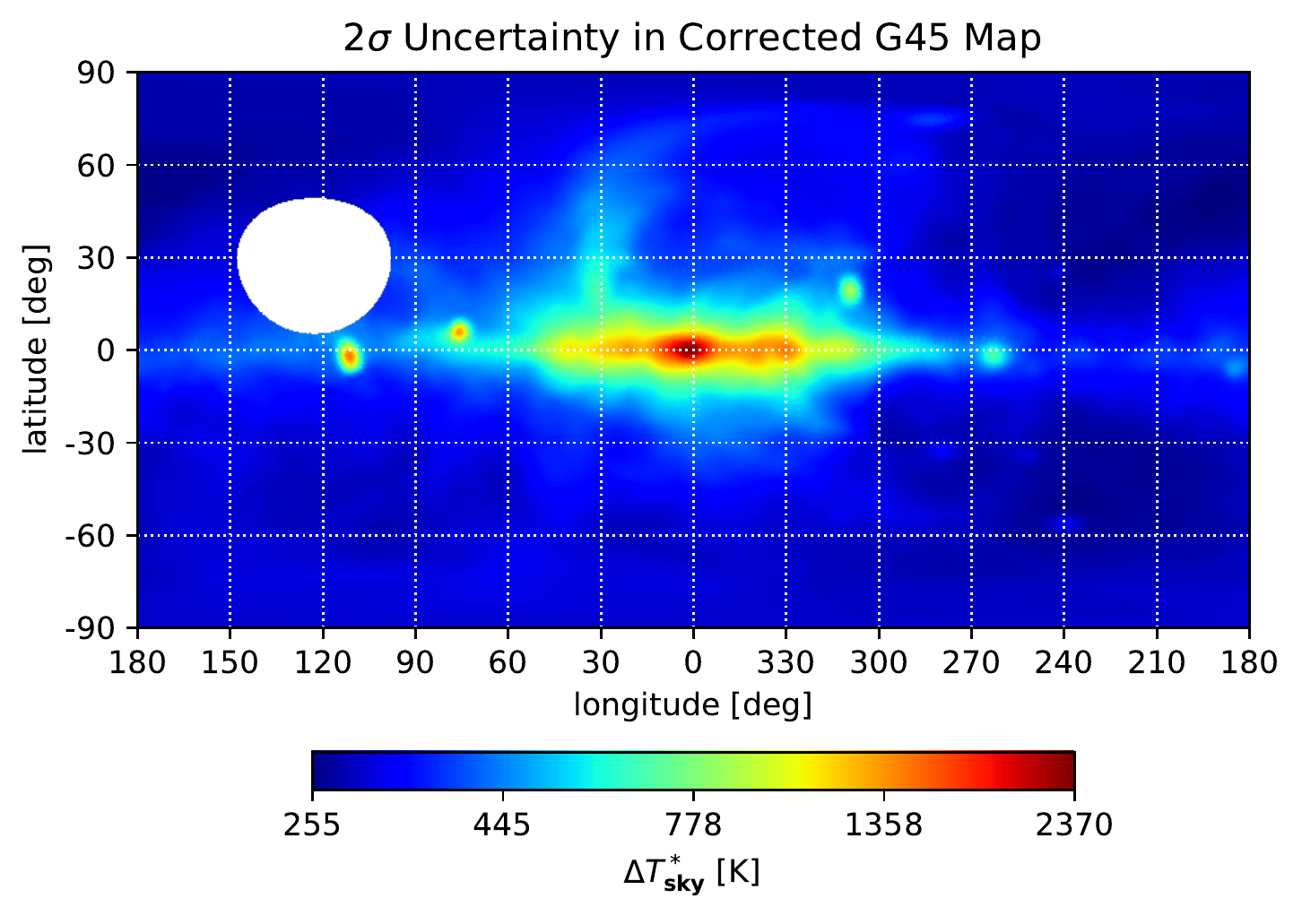}
\includegraphics[width=0.49\textwidth]{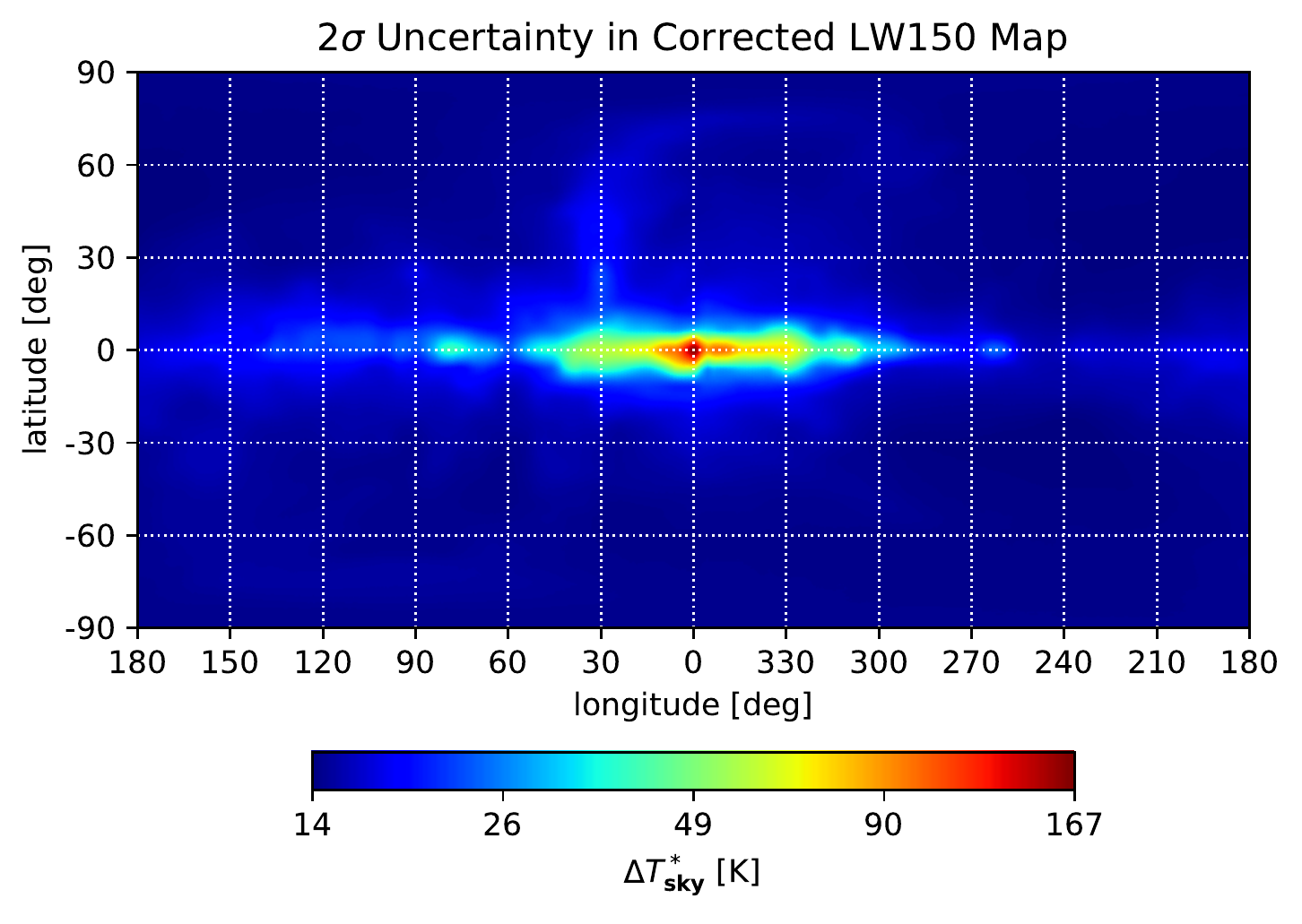}
\caption{(Top) Brightness temperature of the corrected (left) G45 and (right) LW150 maps in Galactic coordinates. (Bottom) $2\sigma$ uncertainties in the corrected maps computed using Equation~\ref{equation_uncertainty_map}. For the G45 map, the percentage uncertainties relative to the corrected temperature are between $3.2\%$ at the Galactic center and $7.5\%$ at high latitudes. For the LW150 map, the percentage uncertainties are between $2.1\%$ at the Galactic center and $9.0\%$ at high latitudes.}
\label{figure_uncertainty_maps}
\end{center}
\end{figure*}

Here, $\hat{T}_{{\rm A},408\rm MHz}$ is the simulated observation of the (original) Haslam map and $\hat{\gamma}$ is our model for the spectral curvature, which is a function of its scaling parameters. The best-fit $\hat{\gamma}$ obtained varies between $-0.0076$ at ${\rm LST}=2$~h and $-0.0332$ at ${\rm LST}=17.3$~h. The result for $\hat{\beta}$ is shown in the right panel of Figure~\ref{figure_consistency_check} using the yellow line. We see that the agreement is now as good as in the case when we corrected the Haslam map assuming zero curvature.

From the previous exercises we take away the following points. The LW150 map was assembled from observations of different regions of the sky at different frequencies with different instruments \citep{landecker1970}. Thus, it is not surprising that the spectral index from simulated observations of the corrected G45 and LW150 maps does not agree very well with the data. Even when we do not correct the Haslam map, the spectral index using the corrected G45 map and the Haslam map agrees better with the data at LST$\sim 22$ to $10$~h than when using the G45 and LW150 maps. A much better agreement can be obtained by either applying corrections to the scale and zero-level of the Haslam map, or by incorporating into the sky model an LST-dependent spectral curvature. We have shown these two possibilities in isolation but in reality expect for both aspects to be necessary in the model. The very strong degeneracy between these two aspects and the lack of precise knowledge of the spectral curvature between $45$ and $408$~MHz prevents us from determining robust corrections to the Haslam map from our low-frequency radio data.

\section{Uncertainty in Corrected Maps}
We now transfer the uncertainties in the scale and zero-level corrections to the actual brightness temperature of the corrected maps. The corrected map brightness temperature is given by $T^*_{\rm sky} = k_1 \cdot T_{\rm sky} + k_2$ and we estimate its $2\sigma$ uncertainty as follows:

\begin{equation}
(\Delta T^*_{\rm sky})^2 = (\Delta k_1 \cdot T_{\rm sky})^2+ (\Delta k_2)^2 + (2 \cdot T_{\rm sky} \cdot \Delta k_{1,2}) + (2\cdot \Delta T_{\rm RMS})^2.
\label{equation_uncertainty_map}
\end{equation}

\noindent The first three terms represent the projection of the correction uncertainties onto the map temperature. $\Delta k_1$ ($\Delta k_2$) is the $2\sigma$ uncertainty on $k_1$ ($k_2$) from Section~\ref{section_systematics} and $\Delta k_{1,2}$ is their covariance, which we compute as the sum of the statistical covariance from the nominal parameter fits and the systematic covariance calculated as

\begin{equation}
\frac{\sum_{i=1}^{N_{\rm syst}} (k_{1i}-\bar{k}_1)(k_{2i}-\bar{k}_2)}{N_{\rm syst}-1}.
\end{equation}

Here, $N_{\rm syst}$ is the number of estimates from the alternative data calibrations and simulations used to determine the systematic uncertainties in Section~\ref{section_systematics}, and the overline represents averaging across estimates. The fourth term in Equation~\ref{equation_uncertainty_map}, $\Delta T_{\rm RMS}$, is an estimate for the errors in the map on angular scales smaller than the monopole. We compute it as the RMS of the difference across LST between the corrected simulations and the measured antenna temperatures, which is shown in the bottom panels of Figures~\ref{figure_summary_45MHz} and \ref{figure_summary_150MHz}. The LST-dependent structure in the difference is not captured by the uncertainty in the scale and zero-level map corrections, which are uniform across LST. Although in principle some of this LST-dependent structure could be due to errors in the calibration of our data and not to errors in the maps, we conservatively project it all toward the uncertainty in map temperature. For each map, $\Delta T_{\rm RMS}$ is a single number computed across both datasets. For the G45 (LW150) map, $\Delta T_{\rm RMS}=108$~K ($6.3$~K).

In Figure~\ref{figure_uncertainty_maps} we show the corrected maps and their $2\sigma$ uncertainty. For the corrected G45 map, the $2\sigma$ uncertainty ranges from $255$~K away from the Galactic plane to $2370$~K at the Galactic center. This represents, respectively, $7.5\%$ and $3.2\%$ of the corrected temperature. For the corrected LW150 map, the $2\sigma$ uncertainty ranges from $14.3$~K ($9.0\%$) at high latitudes to $166.5$~K ($2.1\%$) at the Galactic center. These estimates for $\Delta T^*_{\rm sky}$ are most accurate at angular scales equal to or larger than the scales imposed by our antenna beam patterns. However, due to the diffuse nature of the maps it is reasonable to assume that variations of the uncertainty at smaller spatial scales are low compared to our large-scale estimates.

\section{Relevance for Radio Monopole Excess}

The corrections determined here increase the brightness temperature of the G45 and LW150 maps. Initially this might seem to indicate the presence of an even stronger monopole excess than previously considered. However, it is not trivial to determine the impact of the map corrections on the existance or value of the excess, since this is highly dependent on the accuracy of the estimate for the integrated extragalactic contribution \citep[e.g.,][]{gervasi2008, guzman2011, vernstrom2011, condon2012, vasilenko2017} and, especially, of the model assumed for the Galactic contribution \citep{kogut2011, subrahmanyan2013, fornengo2014, dowell2018}. 

In this paper we do not attempt to investigate the existence of a monopole excess. Nonetheless, because an important quantity in studies of the radio monopole is the minimum brightness temperature in the map, which occurs at high Galactic latitudes, here we report for reference the temperature at coordinates $(l,b)=(+190^{\circ},+50^{\circ})$ after map correction. This point is in the main low-temperature region of the sky above the Galactic plane. For the G45 map, the temperature at these coordinates goes from $3326$~K in the original map, with an uncertainty $\gtrsim \pm 333$~K \citep{alvarez1997, maeda1999, guzman2011}, to $3419\pm 255$~K~($2\sigma$) after correction. For the LW150 map, the temperature goes from $148.9$ with uncertainty $\gtrsim \pm 41$~K \citep{landecker1970}, to $166.3\pm 14.3$~K ($2\sigma$) after correction.

\section{Discussion and Summary}
\label{section_discussion}

In this paper we have derived coordinate-independent corrections to the brightness temperature scale and zero-level of the G45 and LW150 maps. This has been possible due to their (almost) full sky coverage, which is necessary for simulating the wide-beam EDGES observations, and because the map frequencies fall within the EDGES frequency range. Other high-quality radio maps either: (1) do not overlap with EDGES in declination coverage, such as the LWA1 maps from \citet{dowell2017} at $40-80$~MHz which only reach down to $\delta=-40^{\circ}$; or (2) are outside our band, such as the $408$-MHz Haslam map, whose correction with EDGES data would require assumptions for the curvature of the spectral index with which it is highly degenerate.

When fitting simulated observations of the G45 map to our measurements at $45$~MHz we derive a correction of $+7.6\%$ to the  scale of the map, with an uncertainty of $3.4\%$~($2\sigma$). While significantly more precise, this correction is within the uncertainty estimate of $10\%-15\%$ suggested for the scale by \citet{alvarez1997} and \citet{maeda1999}. Our correction for the zero-level of the map is $-160$~K, with an uncertainty of $78$~K~($2\sigma$). Although \citet{guzman2011} calculate a zero-level correction of $-544$~K, this is not comparable to our result since it is not a correction to the actual map brightness temperature but rather the residual they obtain after removing from the map their estimates for the Galactic and extragalactic contributions. The uncertainties we report are almost completely systematic in nature. In the analysis, we used a few nights of data from different times of the year but only with the purpose of maximizing the LST coverage; we did not perform integration across days. This small amount of data and a wide LST coverage were sufficient to keep the statistical component of the correction uncertainty orders of magnitude below the total uncertainty. The systematic effects that have the largest impact on our estimate for the scale of the G45 map are uncertainty (a) in our receiver calibration and (b) in our assumptions for the ionospheric attenuation and emission. The effects with the largest impact on the zero-level are: (a) differences in the results when fitting the two $45$-MHz datasets (with dipole azimuth of $+42^{\circ}$ and $+87^{\circ}$) simultaneously versus each dataset separately, (b) uncertainty in beam pointing along right ascension, and (c) uncertainty in our assumptions for tropospheric and ionospheric refraction. With the map corrections applied, the differences between the simulations and measurements at $45$~MHz are reduced from the range $200-1300$~K across LST to within $\pm 300$~K. The RMS of the corrected difference is $101$~K ($115$~K) for the case with dipole azimuth of $+42^{\circ}$ ($+87^{\circ}$). This represents $\approx 2\%$ of the lowest antenna temperature measured across LST and $\approx 0.7\%$ of the highest temperature. 

From comparisons between EDGES measurements at $150$~MHz and simulated observations of the LW150 map, we determined a scale correction to the map of $+11.2\%$ with an uncertainty of $2.3\%$~($2\sigma$), and a zero-level correction of $+0.7$~K with an uncertainty of $6.0$~K~($2\sigma$). The precision of these corrections is also mainly limited by systematic uncertainties, in particular, uncertainty in receiver calibration. At $150$~MHz, tropospheric and ionospheric effects are small and do not have a significant impact on the estimates. When the corrections are applied to the LW150 map, the differences between the simulations and measurements at $150$~MHz are reduced from the range $20-90$~K across LST to $\pm 20$~K. For the High-Band data with dipole azimuth of $-5^{\circ}$, the RMS of the corrected difference is $5.9$~K. This corresponds to $2.3\%$ ($0.7\%$) of the lowest (highest) antenna temperature measured across LST. For the Mid-Band data with dipole azimuth of $+85^{\circ}$, the RMS is $6.5$~K. This is $2.3\%$ ($0.9\%$) of the lowest (highest) temperature measured. \citet{landecker1970} report a scale uncertainty for the LW150 map of $5-7\%$, which is smaller than our correction. On the other hand, our zero-level correction is small and well within the $40$-K uncertainty they estimate for the map.

Our results for the LW150 map differ from equivalent results from the SARAS experiment \citep{patra2015}. SARAS observed the sky from a latitude of $+13.6^{\circ}$ and reported a scale (zero-level) correction of $+5\%$ ($-22.4$~K) with uncertainty of $0.7\%$ ($8$~K). Although the difference between their results and ours could be attributed to unaccounted systematics in the experiments, it could also be the result of the limited LST coverage ($23$~h to $1$~h) of the data used by \citet{patra2015}, or of significant and different errors relative to the monopole corrections of the LW150 map in the regions of the map observed by SARAS and EDGES. This last possibility is particularly likely since SARAS measured the averaged sky temperature weighted toward the northern celestial hemisphere, which in the LW150 map corresponds to original measurements at $150$~MHz for $-25^{\circ}<\delta<25^{\circ}$ and $178$~MHz for $\delta>+25^{\circ}$ \citep{turtle1962}. EDGES, on the other hand, measured the averaged sky temperature weighted toward the southern celestial hemisphere, which in the LW150 map corresponds to original measurements at $150$~MHz for $-25^{\circ}<\delta<25^{\circ}$ and $85$~MHz for $\delta<-25^{\circ}$ \citep{yates1967}. Another zero-level correction to the LW150 map was suggested by \citet{tartari2008} from measurements with the TRIS experiment at $600$~MHz from a latitude of $+42^{\circ}$. They projected their $600$-MHz data to $150$~MHz using an estimate for the spectral index and after comparing them with simulated observations arrived at a correction of $58\pm 39$~K. The TRIS and SARAS estimates have opposite signs and are different from each other with a significance that is noteworthy, while our zero-level estimate is close to the middle point between the two (considering the error bars).

Having discarded significant polarization effects, one possible cause for the differences observed between our data and the corrected simulations is coordinate-dependent errors in the maps, as suggested earlier. In particular, the possibility of significant residual errors in the corrected LW150 map is supported by our crosscheck in which we compared a measurement of the spectral index of the diffuse sky \citep{mozdzen2019} with a simulation computed using the G45 and LW150 maps. Although the simulation agrees better with the measurement when the corrections are applied to the maps, significant differences remain. These differences largely go away when instead of the LW150 map we use the 408-MHz Haslam map in the simulation.

After correcting the brightness temperature of the maps, we estimated their uncertainty by projecting the uncertainties in the correction parameters to the map domain and adding them in quadrature to the RMS of the difference between our antenna temperature measurements and the corrected simulations. We find that the $2\sigma$ uncertainty in the corrected G45 map ranges from $255$~K at high Galactic latitudes to $2370$~K at the Galactic center. This corresponds, respectively, to $7.5\%$ and $3.2\%$ of the corrected map brightness temperature. For the corrected LW150 map, the $2\sigma$ uncertainty ranges from $14.3$~K ($9.0\%$) at high latitudes to $166.5$~K ($2.1\%$) at the Galactic center.

The corrected maps could be used to refine diffuse radio sky models as well as to re-evaluate the existence of a monopole excess. In this regard, a useful reference value is the brightness temperature in a low-intensity region of the map away from the Galactic plane. We report the value at coordinates $(l,b)=(+190^{\circ},+50^{\circ})$, which is a point in the middle of the main low-temperature region above the Galactic plane. We find a corrected temperature of $3419\pm 255$~K~($2\sigma$) in the G45 map and $166.3\pm 14.3$~K~($2\sigma$) in the LW150 map.

We leave for future work fitting independent scale and zero-level parameters to the different smaller original surveys used to produce the G45 and LW150 maps \citep{turtle1962, yates1967, landecker1970, alvarez1997, maeda1999} in order to increase their accuracy. In addition, the MWA \citep{tingay2013} and HERA \citep{deboer2017} telescopes will likely create new diffuse foreground maps between $50$ and $250$~MHz as they continue to search for the redshifted $21$~cm power spectrum.  Located at the same latitude as these instruments, we anticipate using EDGES observations to help provide absolute calibration for these maps.

\acknowledgements
We are greatful to the reviewer for useful suggestions. We also thank Jack Singal, Dale Fixsen, Alan Kogut, Gilbert Holder, Nicolao Fornengo, Benjamin Harms, Jens Chluba, Jonathan Sievers, Adrian Liu, Avery Kim, and Marta Spinelli for useful discussions. We thank Ravi Subrahmanyan and collaborators for providing to us their version of the LW150 map. This work was supported by the NSF through research awards for the Experiment to Detect the Global EoR Signature (AST-0905990, AST-1207761, AST-1609450, and AST-1813850). N.M. was supported by the Future Investigators in NASA Earth and Space Science and Technology (FINESST) cooperative agreement 80NSSC19K1413. EDGES is located at the Murchison Radio-astronomy Observatory. We acknowledge the Wajarri Yamatji people as the traditional owners of the Observatory site. We thank CSIRO for providing site infrastructure and support.

\section*{ORCID \lowercase{i}D\lowercase{s}}
\noindent Raul A. Monsalve \href{https://orcid.org/0000-0002-3287-2327}{https://orcid.org/0000-0002-3287-2327} \\
Alan E. E. Rogers \href{https://orcid.org/0000-0003-1941-7458}{https://orcid.org/0000-0003-1941-7458} \\
Judd D. Bowman \href{https://orcid.org/0000-0002-8475-2036}{https://orcid.org/0000-0002-8475-2036} \\
Nivedita Mahesh \href{https://orcid.org/0000-0003-2560-8023}{https://orcid.org/0000-0003-2560-8023} \\
Steven G. Murray \href{https://orcid.org/0000-0003-3059-3823}{https://orcid.org/0000-0003-3059-3823} \\
Thomas J. Mozdzen \href{https://orcid.org/0000-0003-4689-4997}{https://orcid.org/0000-0003-4689-4997} \\

\appendix

Here we describe the simulated observations in which we apply typical nighttime tropospheric and ionospheric effects to the G45 and LW150 maps under the assumption that they represent the sky brightness temperature as seen from outside the ionosphere. We point the reader to \citet{vedantham2014}, \citet{sokolowski2015}, \citet{datta2016}, and \citet{shen2020} for detailed discussions of these effects in the context of single-antenna sky radio measurements.

\subsection{Tropospheric and Ionospheric Refraction}

For measurements of the sky brightness temperature from the ground with zenith-pointing, wide-beam antennas, the effect of refraction can be described as a stretching of the antenna gain pattern toward lower elevation outside the ionosphere. I.e., the antenna gain at the zenith angle $\theta$ in the reference frame of the antenna (the `apparent' zenith angle) is shifted to $\theta+\delta\theta$ (with $\delta\theta\geq 0$) outside the ionosphere due to refraction. Here, $\delta\theta$ is the refraction angle, which itself is a function of $\theta$. We incorporate refraction into our simulations with Equation~\ref{equation_convolution} of Section~\ref{section_simulations} by evaluating our antenna gain model at $\theta-\delta\theta$ instead of $\theta$, and integrating with respect to $\theta$ on $[0,\pi/2+\delta\theta(\pi/2)]$. This produces the intended effect of extending the field of view of the antenna to below the horizon. We model the total refraction angle as the sum of refraction from the troposphere and the ionosphere, i.e., $\delta\theta=\delta\theta_{\rm trop}+\delta\theta_{\rm ion}$. We discuss these two effects next.

\emph{Tropospheric Refraction}: Tropospheric refraction occurs due to the altitude gradient of the refractive index in the neutral troposphere. We estimate $\delta\theta_{\rm trop}$ using the approximation recommended by the ITU-R \citep{itu2015},

\begin{equation}
\delta \theta_{\rm trop}(\theta) = \frac{1}{a + b\theta + c\theta^2},
\label{equation_refraction_trop}
 \end{equation}

\noindent where, for angles in radians, $a=16,709.51$, $b=-19,066.21$, and $c=5,396.33$. We note that tropospheric refraction is a function of $\theta$ but not of observation frequency. For reference, the tropospheric refraction angle at the apparent horizon ($\theta=\pi/2$) is $0.76^{\circ}$.

\emph{Ionospheric Refraction}: Based on the model in which ionospheric refraction occurs in the F layer, we compute the ionospheric refraction angle following the approximation developed in \citet{bailey1948}:

\begin{equation}
\delta \theta_{\rm ion} (\nu,\theta) = \left[\frac{\Delta h_{\rm F}(R_{\rm E}+h_{\rm F})}{3R_{\rm E}^2}\nu_{\rm p}^2\right]  \frac{1}{\nu^2} \sin\theta \left[\cos^2\theta + \frac{2 h_{\rm F}}{R_{\rm E}}\right]^{-3/2}.
\label{equation_refraction_ion}
\end{equation}

Here, $R_{\rm E}$ is the Earth's radius, $h_{\rm F}=300$~km is the altitude assumed for the F layer midpoint, $\Delta h_{\rm F}=200$~km is the F layer thickness, and $\nu_{\rm p}$ is the plasma frequency, which for an assumed nighttime total electron content (TEC) in the F layer of $5$~TEC units\footnote{One TEC is equal to $10^{16}$ electrons per m$^{2}$.}, is $4.49$~MHz. $\delta\theta_{\rm ion}$ is given in radians and is a function of both, the observation frequency and the apparent zenith angle. The accuracy of this equation decreases toward the apparent horizon and below. For this reason, we use Equation~\ref{equation_refraction_ion} only for $\theta\leq 80^{\circ}$ and perform a polynomial extrapolation for $\theta>80^{\circ}$. Because in our simulated measurements we also incorporate tropospheric refraction, the apparent zenith angle on the inner side of the F layer is $\theta + \delta\theta_{\rm trop}$ instead of $\theta$. Thus, we compute $\delta\theta_{\rm ion}$ by evaluating Equation~\ref{equation_refraction_ion} at $\theta + \delta\theta_{\rm trop}$. For reference, the ionospheric refraction angle beyond the tropospheric refraction at the apparent horizon ($\theta=\pi/2$) is $0.3^{\circ}$ ($0.027^{\circ}$) at $45$~MHz ($150$~MHz).

\subsection{Ionospheric Attenuation and Emission}

The D layer of the ionosphere attenuates incoming radio waves and is also a source of thermal emission. The brightness temperature of the sky seen by the antenna inside the ionosphere, $T_{\rm sky}$, is related to the brightness temperature outside the ionosphere, $T'_{\rm sky}$, by 

\begin{equation}
T_{\rm sky} = f_{\rm atten}\cdot T'_{\rm sky}+ (1-f_{\rm atten})\cdot T_{\rm e},
\label{equation_attenuation_emission}
\end{equation}

\noindent where $f_{\rm atten}$ is the attenuation factor, which is between $0$ for total attenuation and $1$ for no attenuation, and $T_{\rm e}$ is the D-layer electron temperature. We use the approximated expression in \citet{vedantham2014} for the frequency- and coordinate-dependent attenuation factor:

\begin{equation}
f_{\rm atten}(\nu,\theta) = \exp\left[ -\frac{2 \pi v_{\rm p}^2 v_{\rm c}}{c\left(v_{\rm c}^2 + v^2\right)} \Delta s \right],
\label{equation_attenuation1}
\end{equation}

\noindent with

\begin{equation}
\Delta s = \Delta h_{\rm D} \left(1+\frac{h_{\rm D}}{R_{\rm E}} \right)\left(\cos^2\theta + \frac{2h_{\rm D}}{R_{\rm E}}  \right)^{-\frac{1}{2}}.
\label{equation_attenuation2}
\end{equation}

Here, $c$ is the speed of light, $\nu_{\rm c}$ is the electron collision frequency in the D layer for which we use $5$~MHz \citep{nicolet1953, kane1959, setty1972}, and $h_{\rm D}$ and $\Delta h_{\rm D}$ are the altitude and thickness of the D layer for which we use $75$~km and $30$~km, respectively. To calculate the D-layer plasma frequency, $\nu_{\rm p}$, we use a typical nighttime electron density of $10^8$~m$^{-3}$ \citep{hargreaves1992}. Similarly to Equation~\ref{equation_refraction_ion} for the ionospheric refraction, Equation~\ref{equation_attenuation2} has a lower accuracy toward and below the apparent horizon. Therefore, we use a polynomial extrapolation for $\theta>80^{\circ}$. Because the D layer is between the F layer and the troposphere, instead of computing $f_{\rm atten}$ at $\theta$ we do it at $\theta + \delta\theta_{\rm trop}$. For reference, at $45$~MHz $f_{\rm atten}$ is $0.988$ ($0.983$) at $0^{\circ}$ ($45^{\circ}$), and at $150$~MHz it is $0.999$ ($0.998$) at $0^{\circ}$ ($45^{\circ}$).

The second term in Equation~\ref{equation_attenuation_emission} is the ionospheric emission. Because of its dependence on $f_{\rm atten}$, it is also a function of frequency and zenith angle. To compute this term we assume a typical nighttime value of $T_{\rm e}=800$~K \citep{zhang2004,rogers2015}.

\end{document}